# HEALTH DATA IN AN OPEN WORLD

A REPORT ON RE-IDENTIFYING PATIENTS IN THE MBS/PBS DATASET AND THE IMPLICATIONS FOR FUTURE RELEASES OF AUSTRALIAN GOVERNMENT DATA.

Chris Culnane, Benjamin Rubinstein and Vanessa Teague[1],
School of Computing and Information Systems
The University of Melbourne, 18 Dec 2017
{christopher.culnane, benjamin.rubinstein, vjteague}@unimelb.edu.au

## ABSTRACT

**Abstract**: With the aim of informing sound policy about data sharing and privacy, we describe successful re-identification of patients in an Australian de-identified open health dataset. As in prior studies of similar datasets, a few mundane facts often suffice to isolate an individual. Some people can be identified by name based on publicly available information. Decreasing the precision of the unit-record level data, or perturbing it statistically, makes re-identification gradually harder at a substantial cost to utility. We also examine the value of related datasets in improving the accuracy and confidence of re-identification. Our re-identifications were performed on a 10% sample dataset, but a related open Australian dataset allows us to infer with high confidence that some individuals in the sample have been correctly re-identified. Finally, we examine the combination of the open datasets with some commercial datasets that are known to exist but are not in our possession. We show that they would further increase the ease of re-identification.

## INTRODUCTION

In August 2016, pursuing the Australian government's policy of open government data, the federal Department of Health published online the de-identified longitudinal medical billing records of 10% of Australians, about 2.9 million people. For each selected patient, all publicly-reimbursed medical and pharmaceutical bills for the years 1984 to 2014 were included. Suppliers' and patients' IDs were encrypted, though it was obvious which bills belonged to the same person.

In September 2016 we decrypted IDs of suppliers (doctors, midwives *etc*) and informed the department. The dataset was then taken offline. In this paper we show that patients can also be re-identified, without decryption, by linking the unencrypted parts of the record with known information about the individual. Our aim is to inform policy about data sharing and privacy with a scientific demonstration of the ease of re-identification of this kind of data. We notified the Department of Health of these findings in December 2016.

---

[1] All of the sensitive database queries were conducted by V Teague.

Access to high quality, and at times sensitive, data is a modern necessity for many areas of research. The challenge we face is in how to deliver that access, whilst still protecting the privacy of the individuals in the associated datasets. There is a misconception that this is either a solved problem, or an easy problem to solve. Whilst there are a number of proposals (Australian Government Productivity Commission, 2017), they need further research, development, and analysis. One thing is certain: open publication of de-identified data is not a secure solution for sensitive unit-record level data.

Our motivation in this work is to highlight the challenges and demonstrate the surprising ease with which de-identification can fail. Conquering this challenge will require open and transparent discussion and research, in advance of any future releases. This report concludes with some specific alternative suggestions, including the use of differential privacy for published data, and secure, controlled access to sensitive data for researchers.

## SUMMARY OF RESULTS

Our findings replicate those of similar studies of other de-identified datasets:

- A few mundane facts taken together often suffice to isolate an individual.

- Some patients can be identified by name from publicly available information.

- Decreasing the precision of the data, or perturbing it statistically, makes re-identification gradually harder at a substantial cost to utility.

## STRUCTURE OF THIS PAPER

We first examine uniqueness according to basic medical procedures such as childbirth. We show that some individuals are unique given public information, and show also that many patients are unique given a few basic facts such as year of birth and dates of childbirth.

Although the data is only a 10% sample, we can quantify the confidence of re-identifications, which can be high. We use a second dataset of population-wide billing frequencies, which sometimes shows that the person is unique in the whole population.

We then examine uniqueness according to the characteristics of commercial datasets we know of but do not have. We find high uniqueness rates that would allow linking with a commercial pharmaceutical dataset. We also explain that, consistent with the ``Unique in the shopping mall,'' (de Montjoye, Radaelli, Singh, & Pentland, 2015) financial transactions in the dataset are sufficient for easy re-identification by the patient's bank.

## IMPLICATIONS

Although our specific example is the linked MBS/PBS 10% sample, our results also have implications for the secondary uses of other data such as My Health Records. The de-identification methods were bound to fail, because they were trying to achieve two inconsistent aims: protection of individual privacy and publication of detailed individual records. We demonstrate why that is probably not possible for 30 years of medical billing data. We expect similar results to apply to other rich datasets in the government's care, including census data, tax records, mental health records, penal data and centrelink data.

We support the program of making more data more easily available to facilitate research, innovation and sound public policy. However, there is an important technical and procedural problem to solve: there is no good solution for publishing sensitive unit-record level data that protects privacy without substantially degrading the usefulness of the data. Policy should be made with a clear understanding of the technical ease and serious consequences of re-identification.

This contributes to the debate over the relationship between re-identification, uniqueness and confidence. While uniqueness does not imply re-identification, we show through specific examples that uniqueness, given particular data that is known to be held by certain parties, does imply the opportunity for re-identification. Re-identification is made easier by combining multiple datasets, and can be established with very high confidence although the dataset is only a 10% sample.

## THE RE-IDENTIFICATION AMENDMENT

The Australian government announced plans to amend the Privacy Act to criminalise re-identification of published government data, effective immediately, in September 2016 (Brandis, 2016). The proposed criminal offences would apply if "the information was published on the basis that it was de-identified personal information" 16(D) 1(b). The bill has not (yet) passed, though half of the relevant Senate committee recommended that it should, despite noting, "concerns that have been expressed about aspects of this bill by submitters, including the introduction of criminal offences, the reversed burden of proof and the retrospective application of the bill." (Parliament of Australia, 2017).

A dissenting report (by an equal number of Senators) stated, "the bill adopts a punitive approach towards information security researchers and research conducted in the public interest. In contrast, government agencies that publish poorly de-identified information do not face criminal offences and are not held responsible. … The bill discourages research conducted in the public interest as well as open discussion of issues which may have been identified." We agree.

They also argued that, "retrospective provisions offend a fundamental principle in the rule of law." Again we agree. At the time of the initial announcement, a spokesperson was quoted on the ABC saying that "Provisions will be included in the legislation to allow legitimate research to continue," (Ockenden, 2016) which we interpreted to include all legitimate research. However, when the written bill appeared in October, no such exemption was included, though government entities and others designated by the minister could be exempted. This could easily have put researchers in the difficult position of having conducted what they believed was a legal investigation, only to find that disclosing their findings amounted to admitting to having committed a crime.[2]

Algorithms for protecting online security and privacy need careful scrutiny in order to be improved and strengthened. The introduction of legislative amendments that, whether intentionally or not, have a chilling effect on both the research and wider discourse around this topic, risks critically hampering this effort. Whilst open data is not a safe approach for releasing this type of data, open government is the right paradigm for deciding what is.

We hope this paper contributes to a fair, open, scientific and constructive discussion of these important issues.

### Paper outline

There are two ways to identify individuals in a de-identified dataset: decryption and linking. The first section of this report asks how hard it was to recognise and correct the weakness in the encryption of supplier IDs, which we notified the government about in September 2016. We then examine the difficulty of linking attacks, which involve combining other sources of information with the published dataset to re-identify individual patients.

How hard is this re-identification? Unfortunately, it is straightforward for anyone with technical skills about the level of an undergraduate computing degree.

The later sections examine the possibility of large-scale linking with commercial datasets, then the assessment of confidence of re-identifications.

### decrypting supplier IDs

When we read the description of the method of encrypting supplier IDs, it suggested to us the use of pseudorandom number generation, which was insecure in that setting. We could then guess the algorithm, reverse it, and notify the government, who promptly removed the data from the website. At the time we said that there was a risk that someone else would discover the same weakness.

---

[2] Especially researchers at the ANU or private labs. State-based universities are apparently not covered by federal privacy law.

We learned later that the inappropriateness of that encryption scheme had been discussed extensively on Twitter weeks beforehand. (See weakly de-identified picture.) So we were not the first to discover this problem - it was obvious to many other people.

This specific issue is easy to correct, using any standard encryption algorithm such as RSA or AES. Indeed, encryption was not necessary – a randomly chosen unique number for each person would have worked.

## INDIVIDUAL LINKAGE ATTACKS

*"Succinctly put, 'De-identified' data isn't, and the culprit is auxiliary information."*

*--- Cynthia Dwork*

Computer scientists have used linkage attacks to re-identify de-identified data from various sources including telephone metadata (Srivatsa & Hicks, 2012), social network connections (Narayanan, Shi, & Rubinstein, 2011), health data (Sweeney, 2002) and online ratings (Narayanan & Shmatikov, 2008), and found high rates of uniqueness in mobility data (De Montjoye, 2013) and credit card transactions (de Montjoye, Radaelli, Singh, & Pentland, 2015). Linkage attacks work by identifying a "digital fingerprint" in the data, meaning a combination of features that uniquely identifies a person. If two datasets have related records, one person's digital fingerprint should be the same in both. This allows linking of a person's data from the two different datasets – if one dataset has names then the other dataset can be re-identified. This is not necessarily sophisticated: re-identification based on simply linking with online information has also been reported (Barbaro & Zeller Jr, 2010) (Duhigg, 2002).

### EXAMPLE: THE NETFLIX MOVIE RATINGS DATASET

Netflix, a US DVD rental and streaming service, published a de-identified dataset of users' movie ratings. The IMDb movie knowledgebase website listed users' names along with their ratings. Many users rated a similar (or identical) list of movies on both sites. Different people watch different movies: a person's collection of movies served as a fingerprint, uniquely identifying many users. Narayan and Shmatikov (2008) demonstrated that the Netflix dataset could be re-identified by matching it with IMDb movie ratings and the attached name. One user re-identified in this way had rated sensitive movies privately on Netflix, but not on IMDb. The re-identification technique still worked even given perturbations in the data and some errors in the linking assumptions, because a person's film choices are often very different from everyone else's.

We attempt to apply these techniques to the MBS/PBS dataset.

## BACKGROUND ON THE MBS/PBS 10% SAMPLE DATASET

The MBS/PBS dataset contains billing information, including PBS (prescription) and MBS (medical) records for 10% of Australians born in each year. Each patient has an encrypted ID number and a year of birth and gender. Each record attaches a medical event to a patient, including a code identifying the service or prescription, the state the supplier and patient were in, a date, the price paid by the patient and reimbursed by Medicare and, in the case of MBS records, an encrypted supplier ID. Some rare events were removed before publication, and all the dates were perturbed randomly by up to two weeks in an effort to protect privacy. Some MBS/PBS item codes are generic (code 00023 occurs millions of times, indicating a visit to the GP); others are highly specific and sensitive, such as prescriptions that are only for HIV patients or codes for "management of second-trimester labor".

How easy is it to identify individuals in the dataset by linking the unencrypted parts of the record with known information about the person?

We start by investigating whether the sort of health, demographic, or movement data typically available in a person's public profile is sufficient to identify them uniquely if they are in the released dataset. This is the minimum information that could be available to a malicious attacker.

Many people include on their public profiles their year of birth and gender. This typically puts them in a crowd of more than ten thousand in the MBS/PBS 10% sample. When new information is added, however, the set of possible matches shrinks rapidly. Each new item of known information reduces the set of possible matches within the dataset. When the set of possible matches is small enough to inspect manually, the person's privacy is seriously at risk, if that person is in the dataset.

## Searching for myself (V Teague)

There are nearly 3 million people in the dataset, but only 17,310 women share my year of birth. Two of my children were born in Australia, one in 2006 and the other in 2011. Specifying these years of birth gives 59 possible matches, 23 in my home state of Victoria. Knowing their exact birthdates, and incorporating the two-week perturbation of dates, leaves zero records.

This shows that I am not in the dataset, but if I was I would be easily isolated based on no more information than a typical personal Wikipedia page. This is consistent with other privacy analyses in the literature. A person doesn't need to have a single rare event in order to be identifiable - often a collection of ordinary facts suffices.

Of course, re-identification can be mistaken, but these sorts of results give us some idea of how much information is necessary to identify a person with confidence. If we found that, for about 90% of people, two exact dates of childbirth returned zero matches, then we could infer a fairly low probability of mistaken re-identification based on that much information.

## Searching for others

Information about childbirth patterns is both commonly shared and often unique.

The table below counts births in the dataset from a (deliberately obscured) year b to b+4, for mothers born from year m to m+5. These are women who gave birth quite late in life. Everyone in this age range is unique by years of maternal and child birth, without considering state and without requiring it to be her first child. No cell has more than one, even over the 12 months.

| Year of birth (mother\baby) | b | b+1 | b+2 | b+3 | b+4 |
|---|---|---|---|---|---|
| m | 0 | 0 | 0 | 1 | 1 |
| m+1 | 0 | 0 | 1[3] Dec b+2 or Jan b+3 | 1 Dec b+2 or Jan b+3 | 0 |
| m+2 | 0 | 1 July/Aug | 1 July | 0 | 1 |
| m+3 | 1 | 0 | 0 | 0 | 0 |
| m+4 | 1 | 1 | 1 | 0 | 0 |

**FIGURE 2: NUMBERS OF CHILDBIRTHS BY OLDER MOTHERS, 5 YEAR RANGE.**

---

[3] There is only one childbirth for a mother born in m+1. It is recorded on 2nd January, but because of the 2-week perturbation it might have been in December b+2 or January b+3.

If an attacker knew any of these older mothers, and knew their year of birth and the year their child was born, they could be easily re-identified. The opposite end of the maternal age range is similar: very few childbirths are recorded in which the mother is under 18.

## HOW COULD PRIVACY OF THESE RECORDS BE PROTECTED?

The 2-week perturbation of dates makes no impact on data this sparse. Increasing the perturbation to a few more weeks wouldn't make the slightest difference – we would still have a very small number of women this age giving birth in any given year. Similarly, reducing the precision of the year of mother's birth (for example specifying a 5 year age range rather than the particular year) would still leave a very small number of births, which could be distinguished by state.

The next obvious suggestion is to remove unusual individuals completely, or at least to remove unusual events such as giving birth very late or very early in life. However, this is highly problematic. A person's "anonymity set" is the number of records consistent with all known facts about that person. Mothers in the table above have an anonymity set of 1 if their year of birth and year of giving birth are known. What size of anonymity set is acceptable? For large states like NSW and Victoria, there are quite a few women giving birth in their late 40s. Should we demand the same level of anonymity for smaller states such as SA/NT and WA? How much extra information should be necessary to isolate a particular patient? For example, information about at least one other child is likely to produce a unique match in this set. Is that acceptable?

These important questions do not seem to have good answers. Removing these individuals or events would reduce the accuracy of the data and bias or even prevent some possible studies. Leaving them in is a serious risk to privacy, even with a large perturbation of dates.

Removing unusual individuals is a false option anyway, because everyone is unusual given enough information about them. Indeed, for childbirth as for many other facts about people, a small collection of ordinary events is usually enough to isolate that person. In the next section we discuss identification based on knowledge of 2 or 3 dates of childbirth among women who are not particularly old or young.

## UNIQUE CHILDBIRTH PATTERNS

We examined the sample data to understand how much information about a woman's childbirth history was needed to identify her. An individual in the dataset matches someone else if they were born in the same year and if every time they have a child, the dates are within 4 weeks. (This is the furthest that two events truly on the same day could be shifted apart by the 14-day perturbation.) We asked how much information was necessary to limit one person to a very small number of (other) matches. The results are shown in the dark blue bars of

Figure 3. We have already seen that some very old or very young mothers are uniquely identifiable. So are women who have unusually many children. We found that 102,593 women in the dataset have at most 6 possible matches (including themselves). More than 55,000 are unique. [4]

We then asked whether a larger perturbation of dates would improve privacy, and if so by how much. The results are shown in the same figure, for 28 days (red), 60 days (yellow) and 90 days (green). Like other similar studies, we found that larger perturbations improved privacy only a little. Doubling the perturbation to 28 days still leaves 30,000 unique individuals. Even with a 90-day perturbation, more than 9,000 mothers are unique. In the context of pregnancy and childbirth, a 3 month perturbation of dates would seriously affect the accuracy of many scientific studies.

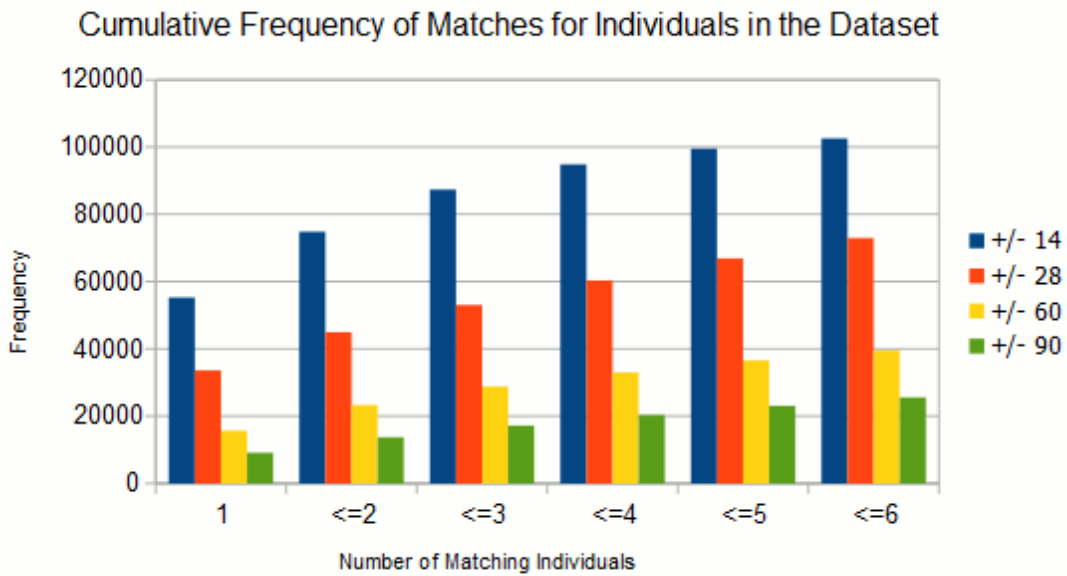

FIGURE 3: MATCHES BY CHILDBIRTH DATES

### Re-identification based on public childbirth information

We read Wikipedia pages and online news stories about famous Australians who are also mothers, recording their years of birth and the birthdates of their children if we could find them. We[5] then entered that information as queries in the database. The first few, including

---

[4] Again, these uniqueness results apply only to women who gave birth in a setting that was billed to medicare. This probably correlates highly with financial resources and hence is relatively easy to guess.

[5] All the sensitive database queries were conducted by V Teague

Prof Gillian Triggs, former Senator Natasha Stott-Despoja, Olympic athlete Cathy Freeman and deputy opposition leader Tanya Plibersek, returned zero matches. Hence we can be fairly confident that those people are not in the sample.[6] This sounds comforting, but really isn't, because it suggests that if they were there they would probably be unique and therefore easily identifiable.

Of 18 queries involving 2 or more childbirth-related events, 13 had no matches in the dataset. One match was rejected as inconsistent with other information about the target person; another was omitted when the online information was found to be inconsistent.

Three queries returned unique matches.

### OTHER HEALTH-RELATED PUBLIC FINGERPRINTS

Many other health-related characteristics could be enough to identify a person. Many other types of people make a great deal of information about their health available on the Internet. We conducted a brief study of professional footballers, noting their injuries and surgeries over the years from online information and translating them into database queries. The results were very similar to those for childbirths. Some people's distinctive histories are clearly absent. For one AFL team captain, a unique record matches the publicly available information about his medical history, year of birth, and interstate movements.

Sometimes a news story gives very precise dates for someone's surgery or hospital admission. In two instances a news story about a relatively common operation on a (former or current) member of a state parliament matches a unique record.

Overall, including the re-identifications from childbirths, sports injuries and single surgeries, we devised 43 queries and found 7 unique matches. This is about 17%, which is consistent with a 10% sample, some randomness, and some probability of a mistaken re-identification based on a coincidental resemblance. We discuss the confidence of these re-identifications below.

### LARGE-SCALE LINKAGE WITH OTHER DATASETS

So far we have used linking information from one-off sources. However, some entities have access to large databases, with names, that could be a source of linking information for many people at once.

---

[6] Many childbirths occur in a public hospital setting that is never billed to medicare and therefore not in the dataset. Hence it is possible that some of those people are in the dataset but their childbirths are not.

Australian privacy law refers to what can "reasonably" be re-identified, but this depends on what other data is available for linking.  Latanya Sweeney purchased the electoral roll database of Cambridge (USA) and used it to re-identify the state governor's medical record (Sweeney, 2002).  The Australian electoral roll is not available for purchase, but is widely distributed.  What other datasets is it reasonable to assume an attacker might be able to access?

## BILLING DATA AND HEALTH INSURERS

One example is the billing data itself: the fees associated with MBS records.  Private health insurers have access to much of their own customers' Medicare claim information.  A simple way to compute the links is to add up the numbers and compare the totals.

The MBS/PBS dataset lists, for each transaction, the amount paid by the patient (or private insurer) and the amount paid by the government.  We computed, for each patient and each year, the government's total payments and the patient's total payments, for MBS and PBS separately, for all the transactions in that year.  This produces four numbers each year, which constitute a fingerprint.  Of the 2.1 million patients with records in 2014, nearly 900,000 received a unique total of paid MBS benefits, while over 900,000 paid a unique contribution.  As we add more years, almost everyone is unique.

Most health insurance companies would be missing some of this data, for example if the item is bulk billed or paid by the patient.  The same approach would work on whatever data they did have.  If the rates of uniqueness were a little smaller on the restricted dataset, totals for different years of data could be used together.

This demonstrates that a private health insurer (for example) could efficiently track the medical records of past customers through the decades of data, or derive extra information they didn't know about from current customers. This would be a clear breach of privacy that would possibly never be reported, even though the data could lead to detrimental decisions for the individual in the future.

The proposed amendments to the Privacy Act criminalizing re-identification would not stop this. Rather than explicitly re-identifying the government dataset, the health insurance company could link it to a de-identified version of their own database, derive some conclusions, delete the MBS/PBS data, and then link their conclusions back to their named clients using the intersecting data.

## PHARMACY RECORDS AND DATA BROKERS

In 2017 the Melbourne hackathon used a de-identified commercial dataset of pharmaceutical records. Although the dataset is not openly available, a metadata file shows that it included unit-record-level longitudinal patient data, and that dates were aggregated into months.

Records could be linked by finding uniqueness in the pharmaceutical part of the government's 10% sample. A few prescriptions suffice. For example in 2003, even without dates, 29,603 people who have two prescriptions are unique in the PBS 10% sample. Figure 4 shows the number of patients with some set of 3 unique prescriptions, for each total number of prescriptions. Overall 147,717 (about 16%) have at least one set of three unique prescriptions in 2003.

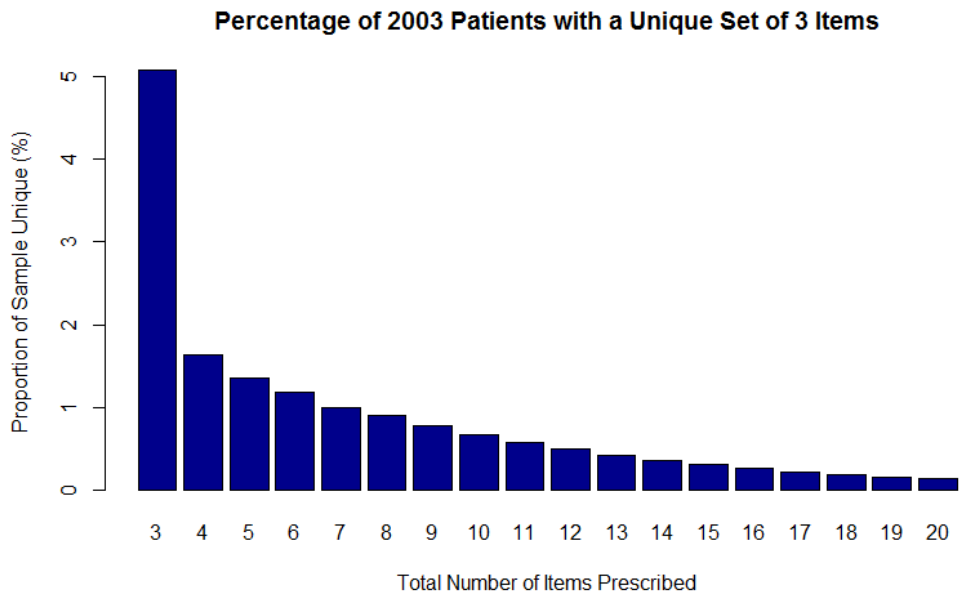

FIGURE 4: NUMBER OF PATIENTS WITH SOME SET OF 3 UNIQUE PRESCRIPTIONS, PER TOTAL NUMBER OF PRESCRIPTIONS.

The extra information in the commercial dataset could be used to increase confidence in matches. Suppose a patient was unique according to 3 matching prescriptions in both datasets. If the date ranges also happened to be consistent, then this would greatly increase the confidence of correct re-identification. The confidence is hard to quantify, but if the dates were chosen completely randomly (the best possible case) then a coincidental match would happen with probability about $(1/6)^3$.

The commercial dataset includes postal code (ZIP code) for both pharmacy and patient. Thus, an accurate link would allow the reattachment of precise location information to the longitudinal 10% sample, which in turn would greatly increase the rates of uniqueness and the confidence of accurate re-identifications.

## PAYMENT DATA AND BANKS OR CREDIT CARD COMPANIES

Credit card records alone can be used to identify an individual. A recent paper on the uniqueness of credit card billing data (de Montjoye, Radaelli, Singh, & Pentland, 2015) found that only 4 transactions were needed to make 90% of people unique, even when prices were removed. Perturbing dates by 15 days improved privacy a little, but 80% of people were still unique based on 10 transactions. The exact numbers might not translate across to the Australian setting, but the implication would: if someone paid for even a handful of doctor's visits or PBS prescriptions with their credit card, then their bank or credit card company would probably have enough information to re-identify their MBS/PBS record.

## LINKING WITH OTHER GOVERNMENT DATASETS

The more individual records the government puts online, the more data can be used in a linkage attack. For example, raw census data includes dates of birth for the entire household.[7] We are not aware of any public ABS microdata that reports this information, but if there was one it could be linked with dates of childbirth in the MBS/PBS dataset. Based on our analysis above, this would provide a unique fingerprint for many women and hence many households. Again some perturbation of dates would make this a little harder, but any reasonable perturbation would still allow some rate of successful linkages. The guarantee that "no personally identifiable data will be released from the census" could therefore depend on what other data is already published and how easily it can be re-identified and linked to census data.

## MALICIOUS ATTACKERS

The information available to well-intentioned researchers represents the absolute minimum information available to someone with criminal intent. The same approach could be used to link information extracted by social engineering, derived from social media by deception, or purchased from leaked or illegally exfiltrated datsets of personal information such as the recent Red Cross leak, which included names and addresses as well as some medical information.

Discoveries about data are often more powerful in combination than alone. The decryption of supplier IDs might not immediately reveal very much about patients, but it could be combined with other data to make a linkage attack much easier. Even without decryption, re-identification of one individual could make others easier, for example if one elite athlete's record reveals which surgeon treats other elite athletes.

---

[7] So does data held by a school or public library.

The threat model here is very strong: a dataset released today might still be scrutinized in the presence of extra information decades into the future.

## SAMPLING, UNIQUENESS AND CONFIDENCE

Several letters in recent issues of *Science* have been devoted to an argument over the meaning of uniqueness results in a sample (Sanchez, Martinez, & Domingo-Ferrer, 2016). One side of the argument emphasizes that

- uniqueness within the sample does not necessarily imply uniqueness within the whole population: a unique match might actually be a coincidental resemblance to someone who is not in the sample, so re-identification cannot be made with confidence, and
- uniqueness does not imply successful re-identification, if you don't know enough points about someone to make them unique.

Although these are both true, they are not as comforting as they sound.

- If the attacker knows the facts that make a person unique, and that person is in the sample, then uniqueness does imply re-identification. If those facts are easily available on the Internet, then uniqueness implies easy re-identification.
- There might be other ways to learn that a particular fingerprint is unique in the whole population. If that fingerprint appears in the sample, then the match is right.
- There might be more information available about a person than is required to make them unique. For example, if 3 facts produce a unique match, and another 2 known facts also match the retrieved record, this would increase confidence.
- Matches of infrequent types could be made with more confidence than matches of more common types. The frequency of occurrences of particular profiles in the sample could be used to estimate the frequency in the whole population.
- Confidence may not be necessary anyway. The identification is accurate if the person is in the dataset.

### EXAMPLE: POPULATION UNIQUENESS BASED ON DHS STATISTICS

The department of human services releases statistics about the rates of MBS billing codes, for the whole population (not just 10%).[8] These are aggregated into patient age ranges of ten years, and reported for each state and each month. Some codes are very uncommon, and are either absent or billed only once in particular months for particular ranges of patient age. Overall, 27% are unique.

---

[8] https://data.gov.au/dataset/medicare-benefits-schedule-mbs-group

For example, former Prime Minister Kevin Rudd had an unusual surgery (an aortic valve replacement) in Brisbane in August 2011. According to the DHS data, this surgery was unique in that month and age range. His record is not in the longitudinal 10% sample dataset, but if it was this could form evidence of a correct re-identification.[9]

This sort of data can also be used to reduce confidence in false identifications. For example, Senator Cory Bernardi was diagnosed with tuberculosis in February or March of 1996. The rate of this disease in Australia is so low that, when we found a man of the right age in the longitudinal 10% sample dataset who had had this test, we were fairly confident that the match was right. However, a careful examination of the DHS data shows that, although in most age ranges and months there are zero or one tuberculosis tests in South Australia, in those particular dates and age ranges there are 4.[10] This turned out to be a false match, and that fact was made more obvious by population-level uniqueness data.

## EXAMPLE: ESTIMATES OF CONFIDENCE FOR RE-IDENTIFICATIONS

Our re-identification results are a concrete example.

Three of our re-identifications are based on multiple childbirth-related events per person. For these, the DHS data does not help estimate confidence.

The other four re-identifications are based upon a single date of surgery, which is unique in the 10% sample within the two-week perturbation, given gender, year of birth and state. We now examine the probability of an incorrect match.

One newspaper article describes a particular (named) patient as the oldest person to have had that surgery in the state. There are two possible item codes, one for the simple version and one for the complex version of this surgery. The DHS data lists one of each in the 10-year age range for the right month. The 10% sample has one consistent with the story, with the "complex" code. It happens to be listed on the 16th of the month, which could not have originated in a different month. If we assume that the oldest person earns the "complex" code then this is conclusive evidence of correct re-identification. If the alternative (simply-coded) person is born in a different year then the re-identification must be right. Without either assumption, there is a 50% probability that it is a false match with the alternative person.

---

[9] This is complicated by the two-week perturbation. Although there were no other patients in August or September, there was one in July whose dates could have been perturbed to form a false positive. There would need to be some evidence against this possibility, *e.g.* if that person was also in the dataset. If not, there would be a 50% chance of accurate re-identification without other information.

[10] This may not be a coincidence: tuberculosis is highly infectious and when one person tests positive their friends are often tested too. Perhaps the 3 other South Australian men of the same age were friends of Sen Bernardi.

Three more re-identifications are by uniqueness based on age, gender, state and dates of surgery. In each case, the DHS data suggest a small total population within the right ten-year age range, but not small enough to be confident of population-wide uniqueness.

In the first case, the surgery occurred in May. DHS data prove this was the only case for that age, gender, state and month. One record in the 10% sample is consistent – it has a date of May 10. However, the DHS data also lists two of the same surgeries in April, so it is possible that a false match occurred if one of those was perturbed to the 10th of May. At most, the probability of a mistaken re-identification is 2/3.

It is tempting to assume that the other two occurrences are randomly chosen among the ten-year age range and over days in April. This would produce a very much lower estimate of the probability of a mistaken re-identification. However, this reasoning is not valid: the two alternatives may for various reasons be the same age as our target, or have had surgery at the end of April. (Recall the example of tuberculosis testing above.)

Analysis for the other two re-identifications yields probabilities of at most 9/10 and 4/5 for a mistaken re-identification. Again a random model of age and date distributions would produce a much lower probability.

There are 91,522 items in the DHS dataset that are unique by month, state and age range and have no matching item in either the prior or following month. Of these, 4919 have a consistent record in the 10% sample with a year of birth guaranteed to put the patient into the right age range. If an attacker knew those patients, this would be evidence of 100% confident re-identification.[11]

More precise dates can be inferred for records linked across the two datasets. If a record is linked, the possible dates for the service are the intersection of those for each dataset alone. For example, an event listed in the DHS data in April but perturbed to May 14 in the 10% sample must have occurred on April 30.

If we had analyzed uniqueness and attempted re-identification over the whole population rather than a 10% sample, we would have a list of perhaps 30 or 40 very confident re-identifications.  Instead we have a list of seven, of which 3 or 4 are likely to be right.  For the 90% of people who are not in the sample, this is a huge gain.  For the 10% who are, the incomplete confidence has little benefit.

---

[11] Unexpectedly, 64 have 2 or 3 matches. We suspect this reflects accidental omissions from the DHS data.

## DISCUSSION: WHERE TO FROM HERE?

Taking advantage of the benefits of big data without seriously compromising privacy is one of the most difficult engineering challenges of our time.  It makes no sense for the government to insist on one solution to this problem – the open publication of de-identified data – despite conclusive evidence that this solution does not work.

The MBS/PBS 10% sample dataset release was motivated by convincing arguments about the utility of that data for medical research that saves lives.  We strongly support this sort of research, and the general aim of informing public policy and inspiring innovation with scientific analysis of data.  The question is how to engineer that without destroying privacy.

Our re-identification of some people in the MBS/PBS 10% sample dataset is not an isolated case, but a replication of numerous other results in which similar techniques have been shown to work on other datasets.  Re-identification will only become easier as more information is released.  The combination of attributes that could form fingerprints is difficult to predict and safeguard against, which is why privacy criteria like k-anonymity (Sweeney, 2002) are inherently flawed (Machanavajjhala, Kifer, Gehrke, & Venkitasubramaniam, 2007) (Li, Li, & Venkatasubramanian, 2007).  It is very unlikely that even the most well-informed and well-intentioned set of guidelines on de-identification can guarantee privacy protections appropriate for sensitive data such as the MBS/PBS 10% sample while retaining the usefulness of the data.

A well-intentioned official might carefully follow a de-identification process, but some individuals might still be "reasonably identifiable".  If compliance with de-identification guidelines is wrongly assumed to imply proper mathematical protections of privacy, this represents a serious loophole in privacy law, not just for government but also for private companies.

Of course it is possible to remove information from records until nothing meaningful can be derived about individual people.  The question is whether this approach solves any of the problems that open data is intended to solve, or whether the manipulation necessary to protect privacy destroys the research value of the data.  If scientists still need to apply for the unperturbed or unaggregated version, then we haven't really solved the problem of data access.

## SOME SPECIFIC QUESTIONS AND SUGGESTIONS

There are many alternatives to full open release of de-identified data. It is better to specify a good process for encouraging research and fact-based decisions about privacy protection and data sharing, than to commit in advance to one particular solution that probably doesn't work.

Different solutions might solve different problems. It is important to ask *exactly what problem are we trying to solve?*

Open public release of data about government is a great idea. Public transport data, expenditure, and other non-sensitive data should be publicly released.

Data about people should be much more carefully considered. It is important to ask why the data is being released, who needs to see it, what they are allowed to do with it, and what the implications of a breach would be. Here is a short list of possible solutions. All of them still include some risk of information leakage, and some inconvenience to researchers. Different approaches address different problems and have different risks - evaluating the tradeoffs would require input from the scientists who use this data.

- Basic de-identification, such as removing names, could be combined with other methods of securing the data.
- Access to data could be controlled either physically or electronically. Researchers could sign up and have their access controlled by a license with penalties for misuse or leakage.
- The Australian Productivity Commission's *draft report on Data Availability and Use* recommended giving de-identified data to "trusted users." These could be required to meet standards for data security.
- Researchers could apply for only those parts of a dataset that they actually need to do their particular research.
- Narayan and Shmatikov (2010) suggest *"An interactive, query-based approach is generally superior from the privacy perspective to the 'release-and-forget' approach."*
- Aggregate data rather than individual records could be released publicly. Frameworks such as Differential Privacy (Dwork, 2008) could be used to assess what aggregates are privacy-preserving.
- In some cases cryptography can be used to make inferences from hidden information, for example to compute totals without decrypting individual records, or to link records from different datasets without decrypting individual IDs. However, this doesn't guarantee that the results of the linking or tallying will protect privacy.
- Expressive consent models such as Dynamic Consent (Kaye, 2015) could allow patients much greater control and visibility over their data.

- Portals such as data.gov.au could list all the available datasets, including sets not published openly on the website.  This could include instructions and requirements for access to data that was available under stricter conditions.

Australia can learn from overseas examinations of the same issues.  The European Union recently released guidelines for the release of public datasets (https://www.europeandataportal.eu/en/content/how-address-privacy-concerns-when-opening-data).  Step 1 is "Understand the data. Consider potential use cases, the value of the data and potential risks."  A de-identification algorithm might be suitable for a scientist working under a legal responsibility to keep the data private, but not suitable for public release of the dataset.

A US presidential commission on cybersecurity received a number of submissions on privacy and data sharing.  An MIT submission (Pentland, Shrier, Hardjono, & Wladawsky-Berger, 2016) emphasised "Open Algorithms" and "Permissible Use."  Open algorithms means that details about the methods and processes should be available for public scrutiny; permissible use emphasises the consent or expectations of the people whose data is shared.

The United Nations Special Rapporteur on the right to privacy recently released a draft report on privacy, big data and open data: http://www.ohchr.org/Documents/Issues/Privacy/A-72-43103_EN.docx

Privacy protection is a subfield of computer science with a peer-reviewed literature that can underpin good decisions about sensitive data.  Privacy should be designed into these processes by people who understand this science.

Publishing details about the de-identification techniques for the MBS/PBS data was the right thing to do because it allowed the problems to be detected.  This approach should be retained and extended: before data is released, details about the process should be published so that they can be examined by privacy experts and the public.  Then the government can make good decisions based on full information from free and open research and public comment.

## Conclusion

Recent government policy relies on secure de-identification of detailed personal data, but secure de-identification of rich data is probably not possible without substantially degrading the data (Ohm, 2010).  This report shows that some MBS/PBS records can be re-identified, thus adding another example to a long list of unsuccessful attempts at de-identification of sensitive datasets.

These failures are not simply a result of choosing a bad method of de-identification.  They reflect the inherent statistical fact that a small number of ordinary points of information is often enough to identify a person.  Perturbing the data or decreasing its precision can improve privacy gradually, at considerable cost to accuracy.  Removing rare individuals or rare events is a false hope because everyone is unique if enough information about them is known.

The proposed amendment to the Privacy Act to criminalise re-identification will not solve these problems.  It will make them harder to detect, understand and avoid.   It inhibits open public analysis and discussion, and hence makes personal data less secure.

There are exciting new ideas for provably privacy-preserving computation on sensitive data, including Differential Privacy, homomorphic encryption and multiparty computation.  These ideas are being put into practice in Australia and overseas.

The Australian government holds vast quantities of information about individual Australians.  It is not really "government data".   It is data about people, entrusted to the government's care.  Data about government should be published openly and freely - not so for sensitive data about people. That should be published only when a clear, public explanation of the encryption and anonymization methods has received enough peer review and public scrutiny to convince everyone that personal information will remain private.  For some datasets, including the MBS/PBS unit-record level data, this is probably not possible.

Making more data available more widely could have many benefits, but the approach needs to be re-engineered with a better understanding of the risks.  Australia really can become a leader in the data sciences by encouraging free and open research in privacy-preserving technologies for data sharing.  Understanding which ideas don't work is a first step in innovation towards techniques that do.